\documentclass[amsmath,12pt,amssymb,preprint,prd,aps,nofootinbib]{revtex4}
\usepackage{amsfonts} 
\usepackage{graphicx} 
\usepackage{epsfig}
\usepackage{multirow}
\usepackage{bm}

\begin{document}
\title{In-medium properties of light vector and axial-vector mesons in magnetized nuclear matter: effects of Dirac sea}

\author{Pallabi Parui}
\email{pallabiparui123@gmail.com}
\author{Amruta Mishra}
\email{amruta@physics.iitd.ac.in}  
\affiliation{Department of Physics, 
Indian Institute of Technology, Delhi, Hauz Khas, New Delhi - 110016}
\begin{abstract}
The in-medium masses of the light vector, $\rho^{0, \pm}$, $\omega$ and the light axial-vector, $A_1^{0, \pm}$ mesons, are studied in the magnetized nuclear matter, accounting for the effects of the Dirac sea. The in-medium partial decay widths for the $A_1\rightarrow \rho \pi$ channels, are studied from the in-medium masses of the initial and the final state particles, by applying a phenomenological Lagrangian to account for the $A_1\rho\pi$ interaction vertices. The masses calculated within the QCD sum rule framework, are obtained in terms of the light quark ($\sim \langle \bar{q}q \rangle$) and the scalar gluon condensates ($\sim \langle G^2 \rangle$), as well as the light four-quark condensate ($\sim \langle \bar{q}q\rangle^2 $). The condensates are calculated within the chiral $SU(3)$ model in terms of the medium modified scalar fields. The effects of magnetic fields are incorporated through the Landau energy levels of protons, anomalous magnetic moments (AMMs) of the nucleons in the nuclear matter, in addition to the magnetized Dirac sea contribution, within the chiral effective model. The enhancement (reduction) of the light quark condensates with magnetic field, is called (inverse) magnetic catalysis. The effects of magnetic fields on the in-medium hadronic decay widths of the $A_1$ mesons, are observed to be significant through the Dirac sea effect, accounting for the AMMs of the nucleons. This may affect the light mesons production in the non-central, heavy-ion collision experiments, where estimated magnetic field is very large.     
\end{abstract}
\maketitle
\section{Introduction}
\label{sec1}
The study of the in-medium hadronic properties (masses, decay widths etc.), under extreme conditions of density and/or temperature, has become a very important topic of research in the strong interaction physics. The study is of great relevance in the ultra relativistic, heavy-ion collision experiments and in the investigation of the size, equation of states of some astrophysical objects, for e.g., neutron stars. In the peripheral, ultra relativistic, heavy-ion collision experiments, huge magnetic fields have been estimated, of the order of $eB\approx 2m_{\pi}^2$ at RHIC in BNL and $eB\approx 15m_{\pi}^2$ at LHC in CERN \cite{kharzeev}-\cite{tuchin}. Thus, the in-medium study of hadrons, in the presence of an external, strong magnetic field background, attracts a lot of research interest in this area. The large number difference between the neutrons and protons of the heavy colliding nuclei, leads to the incorporation of the effects of isospin asymmetry on the study of hadronic properties.\\
The light axial-vector meson, $A_1$ with isospin $I=1$ and the quantum numbers, $J^{PC}=1^{++}$, is considered to be the chiral partner of the much lighter vector meson, $\rho$ with $I^G(J^{PC})=1^+(1^{--})$. The in-medium properties of the light vector mesons can affect the low mass dilepton production \cite{rap} in the heavy-ion collision experiments. The heavy leptonic decay of $\tau \rightarrow \nu_{\tau}+X$, makes it possible to determine the coupling of a hadronic state, $X$ to the axial-vector current. Its decay fraction to the three pion states through the intermediate state of $\rho\pi$, with respect to the total width shows that, the dominant decay mode of the $A_1$ meson is the partial ($s$-wave) $\rho\pi$ mode \cite{tao}. The study of the in-medium spectral properties of the $A_1$ meson is important in the context of partial restoration of chiral symmetry and to provide valuable information to further experimental studies of the axial-vector meson state. There are some well known properties of the axial-vector state, for e.g., its current coupling to the $\pi$ meson, $f_{\pi}$, which is defined in the usual way of $\langle0|\bar{u}\gamma_{\mu}\gamma_5d|\pi\rangle=if_{\pi}P_{\mu}$ \cite{B196}. The fundamental symmetry of Quantum chromodynamics (QCD) is the chiral symmetry, which is spontaneously broken at low densities and low temperatures due to the non-zero expectation values of the QCD vacuum condensates. The formation of the quark and gluon condensates in QCD vacuum, leads to the generation of hadron masses. The spontaneous chiral symmetry breaking effect induces mass splittings between the chiral partners in the spectra of light hadrons, for e.g., $\pi$-$\sigma,$ $\rho$-$A_1$. The condensates are expected to change with temperature and/or density, and also with the magnetic field. The values tend to decrease with increasing density and temperature \cite{91, bhatt, borani}. There is an enhancement (diminution) of the chiral condensates with increasing magnetic fields, a phenomenon called (inverse) magnetic catalysis \cite{Shovkovy, elia, kharzeevmc}. In the literature, there are few studies of the effects of (inverse) magnetic catalysis effects on the hadronic properties in nuclear matter environment. 
There have been some quark matter studies, in the context of the Nambu-Jona-Lasinio model \cite{Preis, menezes, ammc, lemmer, guinjl}. The effects of Dirac sea have been studied on the nuclear matter phase transition in the framework of Walecka model and an extended linear sigma model \cite{haber}. An increasing mass of nucleon as a function of the magnetic field is obtained, at zero baryon density and zero anomalous magnetic moments of the nucleons. The effect of magnetic catalysis is observed indirectly through the scalar field dependency of the nucleon mass, $M_N=m_N-g_{\sigma}\sigma$, where the scalar fields are proportional to the light quark condensates, (with $m_N$ the vacuum mass of nucleon and $g_{\sigma}$ is the $\sigma$ meson nucleon coupling constant). In \cite{arghya}, the Dirac sea effects have been incorporated using the weak-field approximation of an interacting fermion propagator, in the evaluation of nucleonic one-loop self energy functions, at finite magnetic field. An increase of the nucleon mass with magnetic field is observed within the Walecka model, at zero density, in addition to some significant contribution from the anomalous magnetic moments of the Dirac sea of nucleons. At finite temperature, accounting for the anomalous magnetic moments of nucleons, an inverse magnetic catalysis effect \cite{balicm} was observed on the critical temperature of the vacuum to nuclear matter phase transition. The opposite behavior is obtained when the nucleons AMMs are taken to be zero \cite{arghya}. At very high temperatures and very high densities, the chiral symmetric phase is expected to be (partially) restored, which can be inferred from its observable consequence on the hadron spectrum. The dilepton data, as measured in the relativistic heavy-ion collisions, provided evidence on the in-medium spectral changes of $\rho$ meson, while there is very little access to the experimental evidence on the in-medium $A_1$ spectral properties. Therefore, to investigate on the chiral symmetry restoration from the in-medium spectral changes of the $\rho$ meson \cite{hohler}, a theoretical investigations of the same is required on the $A_1$ meson spectrum. Towards this aim, sum rules serve as good non-perturbative tool to connect the hadronic spectral properties directly to the QCD vacuum condensates. The medium modifications of the spectral properties (for e.g., mass, decay width) of hadrons thus observed, may be used as good signals to the changes in the QCD vacuum condensates.\\ There have been QCD sum rule studies for the spectral density induced by the axial-vector current \cite{B196, shifman448}. In these works, Borel sum rules have been constructed for the axial-vector current and divergence of the current, which lead to the same mass value for $A_1^0$ at a particular value of the continuum threshold, $s_0$ \cite{ b283}. The spectral functions of the $\rho$ and $A_1$ mesons have been analyzed in vacuum using the hadronic models (constrained by the experimental data and the Weinberg-type sum rule) and have been tested further with QCD sum rules \cite{rapp}. The Weinberg-type sum rules \cite{weinberg} for the $\rho$ and $A_1$ mesons (in the exact chiral limit), have been studied at zero and finite temperature in ref.\cite{49}; thorough discussion were given on the possible relations between the chiral symmetry restoration and the various types of finite temperature sum rules. In ref. \cite{hatsudab394}, finite temperature (T$\neq$ 0) study of $\rho$, $\omega$ and $A_1$ mesons, by using the Borel transformed sum rule, have been performed. The study has shown a drop in the $A_1$ mass with temperature. The temperature effect have been incorporated through the thermal average of the local operators in the operator product expansion, which leads to the non-vanishing values of Lorentz non-scalar operators that would have otherwise been zero at T=0. The contribution of the scalar four-quark condensates are observed to be significant on these meson properties. In ref.\cite{leupold}, the Breit-Wigner parametrization for the $\rho, A_1$ spectral functions have been used and the constraints of QCD sum rules have been investigated on their masses and decay widths, in vacuum and at finite nuclear density. The effect of the pion coupling to the axial current have been considered by adding a $\delta$-function peak at $m_{\pi}$, to the correlator of axial-vector channel. In ref.\cite{kwon}, the parity-mixing ansatz including the finite widths of $\rho, A_1$ spectra have been investigated at finite temperature in the context of finite energy sum rules. A decreasing $A_1$ meson mass with temperature has been observed, which might indicate the expected tendency of $\rho$ -$A_1$ mass degeneracy in the nearby critical temperature, $T_c$ for chiral symmetry restoration. In ref.\cite{100}, the effects of magnetic field have been studied on the in-medium masses of the light vector mesons ($\rho,\ \omega,\ \phi$), in the nuclear matter environment, by using the sum rule method. In-medium masses are obtained by incorporating the medium effects through the scalar fields, calculated within the chiral $SU(3)$ model. In the absence of an external magnetic field, masses have been studied in the strange hadronic matter, using the sum rule approach \cite{91}, in terms of the condensates calculated within the chiral model via the scalar fields. In ref.\cite{100}, magnetic field contributed through the Landau quantization of protons and the anomalous magnetic moments of the nucleons in the Fermi sea of nucleons. In our present study, effects from the magnetized Dirac sea are incorporated, to study the effects of (inverse) magnetic catalysis on the in-medium masses and decay widths of $\rho$, $\omega$ and $A_1$ mesons at finite magnetic field.\\
The present paper is organized as follows. In sec.\ref{sec2}, the chiral effective model is discussed to find the in-medium quark and gluon condensates. In sec.\ref{sec3}, QCD sum rule approach is presented to calculate the masses of the light vector and axial-vector mesons. Sec.\ref{sec4}, describes the phenomenological Lagrangian formulation to find the hadronic decays of the $A_1$ meson. The results of the present investigation are discussed in sec.\ref{sec5}. Finally, sec.\ref{sec6}, summarizes the findings of the present work.  

\section{The chiral effective Lagrangian}
\label{sec2}
In the present investigation, the in-medium masses of the light vector and axial-vector mesons are calculated within the sum rule approach, by determining the QCD condensates in a chiral effective model. The effective chiral model is based on the non-linear realization of chiral $SU(3)_L\times SU(3)_R$ symmetry \cite{coleman, weinberg1, bardeen} and the broken scale-invariance of QCD \cite{papa, 69, zschi}. A scale-invariance breaking logarithmic potential in the scalar dilaton field, $\chi$ is introduced \cite{sech, ellis}, which simulates the gluon condensate of QCD. The Lagrangian density of the chiral $SU(3)$ model can be generalized as \cite{papa}, \\
\begin{equation}
   \mathcal{L}=\mathcal{L}_{kin}+\mathcal{L}_{BM}+\mathcal{L}_{vec}+\mathcal{L}_0+\mathcal{L}_{scale-break}+\mathcal{L}_{SB}+\mathcal{L}_{mag} 
\end{equation}
$\mathcal{L}_{kin}$ is the kinetic energy of the baryons and the mesons degrees of freedom; $\mathcal{L}_{BM}$ represents the baryon-mesons (both spin-0 and spin-1 mesons) interactions; $ \mathcal{L}_{vec}$ contains the quartic self-interactions of the vector mesons and their couplings with the scalar ones; $\mathcal{L}_0$ incorporates the spontaneous chiral symmetry breaking effects via meson-meson interactions; $\mathcal{L}_{scale-break}$ is the scale symmetry breaking logarithmic potential; and $\mathcal{L}_{SB}$ is the explicit symmetry breaking term; finally the effects of the magnetic field on the charged and neutral baryons in the nuclear medium are incorporated through \cite{p97, am98, am981, prakash, brod, wei, mao} 
\begin{equation}
\mathcal{L}_{mag}=-\frac{1}{4}F_{\mu\nu}F^{\mu\nu}-e_i{\bar{\psi}}_i\gamma_\mu A^\mu\psi_i-\frac{1}{4}\kappa_i\mu_N{\bar{\psi}}_i\sigma^{\mu\nu}F_{\mu\nu}\psi_i
\end{equation}
where,  $\psi_i$ is the baryon field operator for $i^{th} ( i = p, n)$ baryon with electric charge $e_i$, in the nuclear matter. The parameters, $\kappa_p (i=p)= 3.5856$ and $\kappa_n (i=n) = -3.8263$, are the gyromagnetic ratio corresponding to the anomalous magnetic moments (AMMs) of the proton and the neutron, respectively \cite{wei, mao}. In the magnetized nuclear medium, the magnetic field contributions are coming through the Landau energy levels of the protons and the anomalous magnetic moments of the nucleons in the Fermi sea. The Dirac sea effects at finite magnetic field, are incorporated through summation of the scalar ($\sigma$, $\zeta$ and $\delta$) and vector ($\rho$ and $\omega$) mesons tadpole diagrams in the evaluation of one-loop self energy functions of the Dirac sea of nucleons, using the weak-field expansion of nucleonic propagators, within the chiral effective model. The effects of AMMs of the Dirac sea of nucleons are also accounted for in this study. The Dirac sea thus contributes to the scalar densities of the nucleons \cite{chrmmc, dmc, bmc}, which implemented in the scalar fields by solving their coupled equations of motion. The meson fields are treated as classical, whereas the nucleons as quantum fields, in the evaluation of the Dirac sea contribution, in our present study. 
 \\ The concept of broken scale-invariance of QCD is incorporated in the chiral model at the tree level, through a logarithmic potential in the scalar dilaton field $\chi$ \cite{sech, ellis}, as  
\begin{equation}
    \mathcal{L}_{scale-break}=-\frac{1}{4}\chi^4\ln\left(\frac{\chi^4}{\chi_0^4}\right) + \frac{d}{3}\chi^4\ln\left(\left(\frac{(\sigma^2-\delta^2)\zeta}{\sigma_0^2\zeta_0}\right)\left(\frac{\chi}{\chi_0}\right)^3\right)
\end{equation}
where, $\sigma$, $\zeta$  are the non-strange, strange scalar isoscalar fields, respectively and $\delta$ is the scalar isovector field; The Lagrangian parameter, $d$ is chosen to be 0.064 \cite{papa, arindam79}. The subscript-$0$ in the fields indicate their respective vacuum expectation values. The scale-invariance breaking phenomena of QCD leads to the trace anomaly of QCD, i.e., non-zero value for the trace of the energy-momentum tensor in QCD, which in the limit of finite light quark masses $m_i (i=u,d,s)$ become \cite{100, cohen}  
\begin{equation}
    T^{\mu}_{\mu}= \sum_{i=u,d,s}m_i\overline{q}_i q_i+ \frac{\beta_{QCD}}{2g}G_{\mu\nu}^a G^{a\mu\nu}
\end{equation}
In Eq.(4), $G_{\mu\nu}^a$ is the gluon field strength tensor of QCD which is simulated in the chiral effective Lagrangian at the tree level through Eq.(3). The trace of the energy momentum tensor within the model can be obtained from the Lagrangian density terms containing the field $\chi$ \cite{91,arvind82, heide}
\begin{equation}
    \langle\theta_{\mu}^{\mu}\rangle = \chi\frac{\partial\mathcal{L}}{\partial\chi} -4\mathcal{L}=-(1-d)\chi^4
\end{equation}
The first term in Eq.(4) corresponds to the explicit chiral symmetry breaking term in QCD 
\begin{equation}
    \mathcal{L}_{SB}^{QCD}=- Tr[diag(m_u\bar{u}u,m_d\bar{d}d,m_s\bar{s}s)]
\end{equation}
which in the chiral $SU(3)$ model under the mean-field approximation is written as \cite{91} 
\begin{equation}
\mathcal{L}_{SB}= Tr\left[ diag\left( -\frac{1}{2}m_\pi^2 f_\pi(\sigma+\delta),  -\frac{1}{2}m_\pi^2 f_\pi(\sigma-\delta), \left(\sqrt{2}m_k^2f_k-\frac{1}{\sqrt{2}}m_\pi^2 f_\pi\right)\zeta\right)\right]
\end{equation}
Comparing Eqs. (6) and (7), the light quark condensates can be related to the scalar fields $\sigma$, $\zeta$ and $\delta$ as 
\begin{equation}
    m_u\langle\bar{u}u\rangle = \frac{1}{2}m_\pi^2 f_\pi(\sigma+\delta)
\end{equation}
\begin{equation}
    m_d\langle\bar{d}d\rangle=\frac{1}{2}m_\pi^2 f_\pi(\sigma-\delta)
\end{equation}
\begin{equation}
    m_s\langle\bar{s}s\rangle = \left(\sqrt{2}m_k^2f_k-\frac{1}{\sqrt{2}}m_\pi^2 f_\pi\right)\zeta
\end{equation}
From Eqs. (4) and (5), one obtains the scalar gluon condensate as 
\begin{equation}
     \sum_{i=u,d,s}m_i\langle\overline{q}_i q_i\rangle -\frac{9}{8} \left\langle \frac{\alpha_{s}}{\pi}G_{\mu\nu}^a G^{a\mu\nu}\right\rangle =-(1-d)\chi^4
\end{equation}
In the above, $\beta_{QCD}(g) = -\frac{N_cg^3}{48\pi^2} (11-\frac{2}{N_c} N_f )=-\frac{9\alpha_sg}{4\pi}$, the QCD $\beta$ function at the one-loop level is used with $N_f=3$ flavors, $N_c=3$ colors of quarks and $\alpha_s=\frac{g^2}{4\pi}$; by using the expressions for $m_i\langle\bar{q_i}q_i\rangle$, $(i=u,d,s)$ from Eqs. (8)-(10) the final expression for the gluon condensate is given by
\begin{equation}
    \left\langle \frac{\alpha_{s}}{\pi}G_{\mu\nu}^a G^{a\mu\nu}\right\rangle =\frac{8}{9}\left[(1-d)\chi^4 + \left( m_\pi^2 f_\pi \sigma +\left(\sqrt{2}m_k^2f_k-\frac{1}{\sqrt{2}}m_\pi^2 f_\pi\right)\zeta\right) \right]
\end{equation}
The coupled equations of motion in the scalar fields $\sigma$, $\zeta$, $\delta$ and $\chi$, as derived from the chiral $SU(3)$ Lagrangian under the mean-field approximation, are solved in strongly magnetized, isospin asymmetric nuclear matter. In these coupled equations of motion \cite{arvind82}, the number and scalar densities of nucleons ($\rho_i, \rho^{s}_i; i=p,n$) incorporate the effects of the Landau energy levels of protons and the anomalous magnetic moments (AMMs) of the nucleons from the magnetized Fermi sea of nucleons \cite{wei, mao}. The scalar densities of the nucleons incorporate contribution from the magnetized Dirac sea, through sum of the tadpole diagrams in the nucleonic self energy calculation, in a background magnetic field, accounting for the nucleons AMMs \cite{arghya}. The (decreasing) increasing values of the scalar fields with increasing magnetic field represent the (inverse) magnetic catalysis, as the scalar fields are proportional to the light quark condensates as given by Eqs.(8)-(10). Thus, the effects of (inverse) magnetic catalysis in terms of the scalar fields of the chiral effective model, are studied on the in-medium masses of $\rho,\ \omega$ and $A_1$ mesons, in the next section, by using QCD sum rule.  

\section{QCD sum rule framework}
\label{sec3} 
 The masses of the charged and charge neutral light axial-vector mesons, $A_1^{\pm}$, $A_1^{0}$, and light vector mesons, $\rho^{\pm},\ \rho^0$ and $\omega$ are calculated using the QCD sum rule approach. The in-medium masses are obtained in terms of the light quark condensates (including the scalar four-quark condensate) and the scalar gluon condensate, calculated from the chiral $SU(3)$ model. The time-ordered current-current correlator is given by \cite{B196, shifman448}
\begin{equation}
     \Pi_{\mu\nu}(q)= i \int d^4x \  e^{iqx}\left< T[J_{\mu}(x), J_{\nu}(0)] \right>.  
\end{equation}
where T denotes the time-ordered product, and the symbol $\langle\rangle$ indicates the in-medium expectation value of the T-ordered product of currents. The quark-bilinears of currents for all three axial-vector states (with $J^P=1^+$) are defined as 
\begin{equation}
    J^{(A_1^{+})}_{\mu} = \bar{d}\gamma_{\mu}\gamma_{5}u; \quad J^{(A_1^{-})}_{\mu} = \bar{u}\gamma_{\mu}\gamma_{5}d; \quad J^{(A_1^0)}_{\mu}=\frac{1}{2}(\bar{u}\gamma_{\mu}\gamma_{5}u-\bar{d}\gamma_{\mu}\gamma_{5}d)
\end{equation}
For the vector currents (with $J^{P}=1^{-}$)  
\begin{equation}
    J^{(\rho^+)}_{\mu}= \bar{d}\gamma_{\mu}u; \quad J^{(\rho^-)}_{\mu} = \bar{u}\gamma_{\mu}d;\quad J^{(\omega,\ \rho^0)}_{\mu}= \frac{1}{2}(\bar{u}\gamma_{\mu}u\pm\bar{d}\gamma_{\mu}d)
\end{equation}
 The correlation function can be written into the following tensor structure \cite{shifman448, leupold, k1, urban}
 \begin{equation}
     \Pi_{\mu\nu}(q) = q_{\mu}q_{\nu} R(q^2) -g_{\mu\nu} K(q^2)
 \end{equation}
 For the conserved vector currents, $K(q^2)=q^2R(q^2)$. As for the non-conserved, axial-vector current, this relation no longer holds, and $R(q^2)$ has contributions from pseudoscalar mesons \cite{B196, k1, shifman448, urban}. In principle, sum rule can be carried out with either $R(q^2)$ or $K(q^2)$ \cite{shifman448}. Further studies in this work are based on $R(q^2)$, following the arguments given in ref.\cite{leupold}, for the axial-vector current. \\ 
 There are two representations of the current-current correlator. On the phenomenological side, $R(q^2)$ is related to its imaginary part via the dispersion relation \cite{91,k1} 
 \begin{equation}
       R_{phen.} (q^2) = \frac{1}{\pi}\int^{\infty}_0 ds\ \frac{ImR^{phen.} (s)}{(s-q^2)}
  \end{equation}
 Here $ImR^{phen.}(s)$, also called the spectral density, is parametrized by the hadronic resonance and the perturbative continuum part. The other representation is to express the real part of the correlation function by the Wilson's operator product expansion (OPE), which in the large space like region $(Q^2=-q^2)>>1$ $GeV^2$ \cite{hatsudab394,leupold,klingl}, is given by     
\begin{equation}
    R_{OPE} (q^2=-Q^2) = \left( -c_0 \ln\left(\frac{Q^2}{\mu^2}\right) + \frac{c_1}{Q^2} + \frac{c_2}{Q^4} + \frac{c_3}{Q^6} + ...\right) 
\end{equation}
  In the present study, the operators up to dimension-6 are considered and the scale $\mu$ has been chosen as 1 GeV \cite{91,hatsudab394, klingl}. The first term in the OPE is a contribution from perturbative QCD. The coefficients $c_i$  $(i=1,2,3)$ of the subsequent terms, contain the QCD non-perturbative effects in terms of the light quark and gluon condensates and some parameters from the QCD Lagrangian. The condensates are affected in presence of a medium. The coefficient, $c_3$ is being associated with the light four-quark condensate and it is different corresponding to the different current quark-bilinears of the charged and neutral mesons with the same $J^{PC}$ quantum numbers. In Eq.(18), the $c_i$ $(i=0,1,2,3)$ coefficients for the light axial-vector meson, $A_1$ are given by \cite{hatsudab394, leupold, k1, F1} 
\begin{equation}
    c_0=\frac{1}{8\pi^2} \left( 1+\frac{\alpha_s}{\pi}\right )  
\end{equation}
\begin{equation}
    c_1=-\frac{3}{8\pi^2}(m_u^2+m_d^2)
\end{equation}
\begin{equation}
    c_2=\frac{1}{24}\langle\frac{\alpha_s}{\pi}G^{\mu\nu}G_{\mu\nu}\rangle - \frac{1}{2}\langle m_u \bar{u}u+m_d\bar{d}d\rangle 
\end{equation}
\begin{multline}
    c_3^{(A_1^0)}=-\frac{\pi}{2}\alpha_s\Bigg[\left\langle\left(\bar{u}\gamma_{\mu}\lambda^a u - \bar{d}\gamma_{\mu}\lambda^ad\right)^2\right\rangle   + \frac{2}{9} \bigg\langle\left(\bar{u}\gamma_{\mu}\lambda^a u + \bar{d}\gamma_{\mu}\lambda^ad\right) \times \\
\left(\sum_{q=u,d,s}\bar{q}\gamma^{\mu}\lambda^a q\right)\bigg\rangle\Bigg]   = \pi \alpha_s \times \frac{88}{81}\kappa_{1}\left(\langle\bar{u}u\rangle^2 + \langle\bar{d}d\rangle^2\right)
\end{multline}
\begin{multline}
    c_3^{(A_1^{\pm})}=-\frac{\pi}{2}\alpha_s\Bigg[2\left\langle\left(\bar{u}\gamma_{\mu}\lambda^a d\right)\left(\bar{d}\gamma_{\mu}\lambda^a u\right)\right\rangle   + \frac{2}{9} \bigg\langle\left(\bar{u}\gamma_{\mu}\lambda^a u + \bar{d}\gamma_{\mu}\lambda^ad\right) \times \\
\left(\sum_{q=u,d,s}\bar{q}\gamma^{\mu}\lambda^a q\right)\bigg\rangle\Bigg]   = \pi\alpha_s\times \kappa_{2} \left[ \frac{16}{81}\left(\langle\bar{u}u\rangle^2 + \langle\bar{d}d\rangle^2\right) + \frac{16}{9}\langle\bar{u}u\rangle\langle\bar{d}d\rangle \right]
\end{multline}
The first three coefficients, $c_i$ $(i=0,1,2)$ are same for the charged and neutral $A_1$ mesons. To simplify the expression for the scalar four-quark condensates, the use of the factorization method \cite{shifman448, b283} is adopted
\begin{equation}
    -\langle(\bar{q_i}\gamma_{\mu}\lambda^a q_j)^2\rangle=\langle(\bar{q_i}\gamma_{\mu}\gamma_{5}\lambda^a q_j)^2\rangle=\delta_{ij}\frac{16}{9}\kappa_j\langle\bar{q_i}q_j\rangle^2
\end{equation}
The parameter $ \kappa_j$ here, introduces deviation from the exact factorization \cite{91} (which is $ \kappa_j=1$ in the vacuum saturation assumption \cite{shifman448, zos}). The value of the running coupling constant is $\alpha_s = 0.35$, at $\mu=1$ GeV scale, as in refs.\cite{91, hatsudab394}.
For the (neutral) vector meson channels of $\rho$ and $\omega$, the Wilson coefficients are taken from refs.\cite{91, 100} in order to calculate the effects of (inverse) magnetic catalysis on their masses. For the charged $\rho$ meson channel, the first three coefficients ($c_i; i=0,1,2$) in the OPE are same as its neutral partner ($\rho_0$). The $c_3$ term is different in the following way
\begin{multline}
    c_3^{\rho^{\pm}}=-\frac{\pi}{2}\alpha_s\Bigg[2\left\langle\left(\bar{u}\gamma_{\mu}\gamma^{5}\lambda^a d\right)\left(\bar{d}\gamma_{\mu}\gamma^{5}\lambda^a u\right)\right\rangle   + \frac{2}{9} \bigg\langle\left(\bar{u}\gamma_{\mu}\lambda^a u + \bar{d}\gamma_{\mu}\lambda^ad\right) \times \\
\left(\sum_{q=u,d,s}\bar{q}\gamma^{\mu}\lambda^a q\right)\bigg\rangle\Bigg]   = \pi\alpha_s\times \kappa_{3} \left[ \frac{16}{81}\left(\langle\bar{u}u\rangle^2 + \langle\bar{d}d\rangle^2\right) - \frac{16}{9}\langle\bar{u}u\rangle\langle\bar{d}d\rangle \right]
\end{multline}
The QCD sum rule, thus connects the hadronic spectral properties in terms of the resonance parameters given by Eq.(17), to the non-perturbative effects of QCD through the quark and gluon condensates in Eq.(18), with $c_i$'s given by Eqs.(19)-(25).\\
In practice, the parametrization of the current-current correlator is given for the energy region of the lowest hadronic resonance, usually there is no model which can be valid for arbitrary high energies. In this purpose, it is desirable to achieve a larger suppression on the high energy part of the hadronic spectral distribution, which is least accessible. Borel transform is applied in this aim. It is defined on an arbitrary function $g$ as \cite{k1} 
\begin{equation}
    \textit{g}\ (Q^2) \ {\xrightarrow{\hat{B}}} \ \widetilde{\textit{g}}\ (M^2)
\end{equation}
\begin{equation}
    \hat{B}:= \lim_{Q^2\to\infty, n\to\infty }  \frac{1}{\Gamma(n)} (-Q^2)^n \left(\frac{d}{dQ^2}\right)^n. 
\end{equation}
with $Q^2/n= M^2$ become fixed and $M$ is called the Borel mass. 
By the application of Borel transform, the phenomenological side of Eq.(17) is connected to the OPE side in Eq.(18) as  
\begin{equation}
   \frac{1}{\pi} \int^{\infty}_0 ds \ e^{-s/M^2} \ Im R^{phen.}(s) =  \left[c_0M^2 + c_1+ \frac{c_2}{M^2} + \frac{c_3}{2M^4} \right]
\end{equation}
The exponential function on the l.h.s enhances the effect of the ground state by suppressing the continuum part at large $s$. The higher dimensional operators of the $R_{OPE}$, on the r.h.s, are suppressed by an additional factor of $1/(n-1)!$; which leads to the better convergence of the operator product expansion.\\
The hadronic spectral density function, $Im R^{phen.} (s)$ separates into a resonance part, $R_{res}(s)$ (for $s\leq s_0$) and a perturbative continuum (for large s) \cite{91,leupold,klingl}
\begin{equation}
   \frac{Im R^{phen.} (s)}{\pi}   = R_{res}(s)\ \Theta(s_0 - s) +  c_0 \ \Theta(s- s_0 )
\end{equation}
Where $s_0$ is the threshold between the low energy resonance region and the high energy continuum part, latter being calculated in perturbative QCD. \\
Next, the finite energy sum rules (FESRs) are derived by inserting Eq.(29) into the l.h.s of Eq.(28). The exponential function then expanded in powers of $\frac{s}{M^2}$ for $s\leq s_0$, and $M>\sqrt{s_0}$. Equating the various powers of $1/M^2$ on both sides of Eq.(28), after this expansion, the finite energy sum rules in vacuum, can be written as 
 \begin{equation}
     \int^{s_0}_0 ds \ R_{res}(s) =  (c_0 s_0 + c_1)
 \end{equation}
 \begin{equation}
     \int^{s_0}_0 ds \ s R_{res}(s) =  \left(\frac{c_0 s_0^2}{2} - c_2\right)
 \end{equation}
 \begin{equation}
 \int^{s_0}_0 ds \ s^2  R_{res}(s) =  \left(\frac{c_0 s_0^3}{3} + c_3\right)
 \end{equation}
  The parametrization of the spectral function, $R_{A}^{phen.} (s)$ used in the present study for the axial-vector channel is 
  \begin{equation}
           R_{A}^{phen.} (s)= F_{A} \delta(s-m_A^2) + c_0\Theta(s-s_0^A) + f_{\pi}^2\delta(s-m_{\pi}^2)
     \end{equation}
Where, $\frac{Im R^{phen.} (s)}{\pi}=R^{phen.}(s)$. In the axial-vector channel, apart from the $A_1$ resonance term, there is a contribution from the pseudoscalar meson resonance, due to its coupling to the axial-vector current \cite{B196, shifman448,leupold}. The simple ansatz for the vector meson channel is given by \cite{91, 100,klingl, hatsudab394}
\begin{equation}
R_{V}^{phen.}(s) = F_{V} \delta(s-m_V^2) + c_0\Theta(s-s_0^V) 
\end{equation}
Inserting the expression (33) into Eqs.(30)-(32), the finite energy sum rules for the axial-vector channel, in the vacuum, are given by
\begin{equation}
    F_A= \left(c_0 s_0^A + c_1 - f_{\pi}^2\right)
\end{equation}    
\begin{equation}
    F_Am_A^2= \left(\frac{c_0 (s^A_0)^2}{2} - c_2 - f_{\pi}^2m_{\pi}^2\right)
\end{equation}
\begin{equation}
    F_A m_A^4 = \left(\frac{c_0 (s_0^A)^3}{3} + c_3^A -f_{\pi}^2m_{\pi}^4\right)
\end{equation}
In the same way, for the vector meson channel in the vacuum, using Eq.(34) into Eqs.(30)-(32), one obtains 
\begin{equation}
    F_V= \left(c_0 s_0^V + c_1 \right)
\end{equation}    
\begin{equation}
    F_Vm_V^2= \left(\frac{c_0 (s^V_0)^2}{2} - c_2 \right)
\end{equation}
\begin{equation}
    F_V m_V^4 = \left(\frac{c_0 (s_0^V)^3}{3} + c_3^V \right)
\end{equation}
In the nuclear medium, for mesons at rest, the meson-nucleon scattering effect is incorporated through the Landau damping term, $\rho_{sc}$, in the spectral function,

\begin{equation}
  \int^{\infty}_0 ds \ e^{-s/M^2} \ \frac{Im R^{phen.}(s)}{\pi}  + \rho_{sc} =  \left[c_0M^2 + c_1+ \frac{c'_2}{M^2} + \frac{c'_3}{2M^4} \right] 
\end{equation}
Where the primed symbols denote the corresponding in-medium quantities. The damping term originates due to the absorption of a space-like meson by an on-shell nucleon \cite{leupold, flower}. The contributions for the vector and axial-vector meson channels are taken to be the same based on their scattering amplitudes with the nucleons \cite{flower, klingl}. Incorporating the scattering term as, $\rho_{sc}=\frac{\rho_B}{4M_N}$ \cite{91, klingl, prc52, flower} in two channels (for both charged and neutral mesons with different $c'_3$), the modified FESRs [Eqs.(35)-(37)], in medium  are given by 
\begin{equation}
    F'_A= \left(c_0 s_0^{'A} + c_1 - f_{\pi}^2 - \frac{\rho_B}{4M_N}\right)
\end{equation}    
\begin{equation}
    F'_Am_A^{'2}= \left(\frac{c_0 (s^{'A}_0)^2}{2} - c'_2 - f_{\pi}^2m_{\pi}^2\right)
\end{equation}
\begin{equation}
    F'_A m_{A}^{'4} = \left(\frac{c_0 (s_0^{'A})^3}{3} + c_3^{'A} -f_{\pi}^2m_{\pi}^4\right)
\end{equation}
The in-medium FESRs for the vector meson channel \cite{91} are thus modified to
\begin{equation}
    F'_V= \left(c_0 s_0^{'V} + c_1 - \frac{\rho_B}{4M_N}\right)
\end{equation}    
\begin{equation}
    F'_Vm_V^{'2}= \left(\frac{c_0 (s^{'V}_0)^2}{2} - c'_2 \right)
\end{equation}
\begin{equation}
    F'_V m_V^{'4} = \left(\frac{c_0 (s_0^{'V})^3}{3} + c_3^{'V} \right)
\end{equation}
The coefficient in the scalar four-quark condensate, $\kappa_j$, for the vector and axial-vector states, are determined by solving their respective vacuum FESRs. Once this factor is determined, the in-medium resonance parameters of the spectral function, i.e., mass, $m'$, the strength, $F'$ and the threshold energy, $s'_0$ can be obtained by solving the in-medium FESRs. 

\section{In-medium Decay widths of $A_1$ meson}
\label{sec4}
In this section, the hadronic decay modes of the light axial-vector meson, $A_1$ going to a light vector meson, $\rho$ and a pseudoscalar meson, $\pi$ are investigated. It is an observed decay mode of the $A_1$ meson, as given in the particle data group \cite{pdg}. A phenomenological Lagrangian is introduced for the $av\phi$ interaction (here, $a,v, \phi$ denote the axial-vector, vector and pseudoscalar meson fields, respectively)
\begin{equation}
      \mathcal L_{{av\phi}}= i \tilde{f}  \langle a_{\mu\nu}[v^{\mu\nu}, \phi]\rangle.
 \end{equation}
Where, the $SU(3)$ matrices of the mesons are considered as given in ref.\cite{roca70}; The symbol $<>,$ denotes the trace of the  product of matrices and $i$ in front of $\tilde{f}$ is to make the Lagrangian hermitian. The spin-1 meson fields in Eq.(48) are treated as anti-symmetric tensor fields, $a_{\mu\nu}$ and $v_{\mu\nu}$ which transform under a non-linear realization of (local) chiral symmetry, $G=SU(3)_L\times SU(3)_R $ as \cite{eckerb321},
\begin{equation}
    X_{\mu\nu} \ {\xrightarrow{G}}\ h(\phi') X_{\mu\nu} h(\phi')^{\dagger}, \quad X_{\mu\nu}=a_{\mu\nu}, v_{\mu\nu}
\end{equation}
The chiral $SU(3)_L\times SU(3)_R $ symmetry group is spontaneously broken down to $SU(3)_V$ group. A non-linear realization of $G$ \cite{coleman}, is defined on the elements $u(\phi')$ of the coset space $SU(3)_L\times SU(3)_R/ SU(3)_V $ as 
\begin{equation}
    u(\phi') \ {\xrightarrow{G}}\ g_R u(\phi') h(\phi')^{\dagger}=h(\phi') u(\phi') g_L^{\dagger};\ \ g_{R, L}\in SU(3)_{R,L},\  h(\phi')\in SU(3)_V
\end{equation}
where $\phi'$ are the Goldstone boson fields and $\phi=\frac{1}{\sqrt{2}}\sum_{i=1}^8\lambda_i\phi_i'$. The elements of the coset space are defined as, $u(\phi')=exp(-\frac{i}{\sqrt{2}}\frac{\phi}{f})$. The vector, axial-vector and pseudoscalar meson fields transform as octets under $SU(3)_V$. If the $SU(3)$ multiplets are denoted by $S$ then, the non-linear realization of $G$ leads to the following transformations 
\begin{equation}
   S \ {\xrightarrow{G}}\ h(\phi') S h(\phi')^{\dagger}, \quad S= a, v, \phi 
\end{equation}
The transformations, Eqs.(49)-(51) preserve the symmetry of the Lagrangian. Thus, a phenomenological Lagrangian is used for the $A_1\rho\pi$ vertices (along with the other vertices in the octet of mesons), to find the decay widths of $A_1\rightarrow \rho\pi$ channels \cite{roca70}. To calculate the hadronic decay widths for various $(A_1\rightarrow \rho \pi)$ channels, the amplitudes are derived from the above Lagrangian. 
The various strong decay modes of axial-vector mesons going to vector and pseudoscalar mesons are noted in refs.\cite{pdg}-\cite{roca70}; some of the decays are only observed, no experimental values are provided for the corresponding decay widths. The possible decay channels of $J^{PC}=1^{++}$ and $J^{PC}=1^{+-}$ family of axial-vector mesons have been considered with their respective branching ratio, consistent with the particle data group information in the global fit of the coupling parameters (here only $\tilde{f}$ is needed) (table.(IV) of \cite{roca70}). These fittings are described in detail in that reference using both $J^{PC}=1^{+-}$ and $J^{PC}=1^{++}$ multiplets of axial-vector mesons, namely the class of $B_1$ and $A_1$ mesons. The values of the fitted parameter, $\tilde{f}$ are used here to find the in-medium partial decay widths of the $A_1$ meson within the nuclear medium in presence of a background magnetic field, including the contribution of Dirac sea on their masses. For tree level calculations, one can use the form, $(\partial_{\mu}X_{\nu}-\partial_{\nu}X_{\mu})$ instead of the tensor, $X_{\mu\nu}$ for $X = a, v$. Normalization of these fields are given by \cite{eckerb321}
 \begin{equation}
     \langle 0 | X_{\mu\nu} | X; p, \epsilon \rangle = \frac{i}{m_X}[p_{\mu}\epsilon_{\nu}(X)-p_{\nu}\epsilon_{\mu}(X)].
 \end{equation}
  $m_X$ denotes the mass of the $X$ meson. The expression for the $A_1 \rightarrow \rho\pi$ decay width can be written as
 \begin{equation}
     \Gamma_{A_1 \rightarrow \rho\pi} = \frac{q}{8\pi m_{A_1}^2} \overline{|\mathcal{M}|^2} 
 \end{equation}
 Where, $q$ is the momentum of the final state particles in the rest frame of the axial-vector meson, given by 
 \begin{equation}
     q(m_{A_1}, m_{\rho}, m_{\pi}) = \frac{1}{2 m_{A_1}} \Big([m_{A_1}^2-(m_{\rho}+ m_{\pi})^2] \times [m_{A_1}^2-(m_{\rho}- m_{\pi})^2] \Big)^{1/2}
 \end{equation}
 The interaction terms corresponding to the $A_1\rho\pi$ vertices as given by Lagrangian (48), are
 \begin{equation}
    \mathcal{L}_{A_1\rho\pi} = \sqrt{2}i\tilde{f} \Big[ A_1^0\left(\rho^-\pi^+ -\rho^+\pi^-\right) + A_1^+\left(\rho^+\pi^0 - \rho^0\pi^+ \right) + A_1^-\left(\rho^0\pi^- -\rho^-\pi^0 \right) \Big] 
 \end{equation}
 These terms result from the parity and charge conjugation conservation of the Lagrangian density.
 The amplitude, obtained from the interaction vertices, becomes 
 \begin{equation}
     \mathcal{M} = \frac{-2\lambda_{av\phi}}{m_{A_1}m_{\rho}} \left(p'.p \quad\epsilon'.\epsilon - \epsilon'.p\quad \epsilon.p'\right)
 \end{equation}
 with $p'$, $ \epsilon'$ and $p$, $\epsilon$ are the four-momenta and polarization vectors of the $A_1$ and $\rho$ mesons respectively. Using the tree level form of the tensor fields, as given by equation (52), one finds the gauge invariant amplitude for the decay process. The coefficients, $\lambda_{av\phi}=\sqrt{2}i \tilde{f}$ (with the appropriate sign). Therefore, the decay width is given by Eq.(53) by using Eqs.((54)-(56)) and the invariant relations between four-momenta, $p^2=m_{\rho}^2$, $p'^2=m_{A_1}^2$ and $2p.p'=(m_{A_1}^2+m_{\rho}^2-m_{\pi}^2)$, 
 \begin{equation}
     \Gamma_{A_1 \rightarrow \rho \pi} =  \frac{|\lambda_{av\phi}|^2}{2\pi m_{A_1}^2}q\left(1+\frac{2q^2}{3m_{\rho}^2}\right)
 \end{equation}

 \section{Results and Discussions}
 \label{sec5}
In the present work, the in-medium masses of the light vector mesons, $\rho^{0, \pm}$, $\omega$ and the light axial-vector mesons, $A_1^{0, \pm}$ are investigated in the asymmetric nuclear medium, in presence of an external magnetic field, accounting for the Dirac sea effects. The masses are computed within the QCD sum rule approach, by using the medium modified light quark and scalar gluon condensates and the parameters of QCD Lagrangian (the current quark mass, $m_u$, $m_d$, and the QCD running coupling constant, $\alpha_s$). The medium modifications of the condensates are obtained from the medium modified scalar fields, $\sigma$, $\zeta$, $\delta$ and $\chi$, within the chiral $SU(3)$ model [as given by Eqs.\textbf{(8)-(10)} and Eq.\textbf{(12)}]. In the chiral $SU(3)_L\times SU(3)_R$ model, the meson fields are considered to be classical fields under the mean field approximations, leaving the contributions of the scalar fields and time-like component of the vector fields. The nucleons are treated as the quantum fields in the study of the Dirac sea effects. Thus, the coupled equations of motion in the scalar fields are solved by considering the effects of the baryon density $\rho_B$, isospin asymmetry $\eta  \big(=\frac{\rho_n - \rho_p}{2\rho_B}\ \big)$, and an external magnetic field $|eB|$, through the number ($\rho_{p,n}$) and scalar densities ($\rho^s_{p,n}$) of the protons and neutrons in the magnetized nuclear medium. At finite magnetic field, there are contributions from the Landau energy levels of the charged nucleons (i.e., protons) in the magnetic nuclear matter. There is another effect of $|eB|$ accounting for the anomalous magnetic moments (AMMs) of the nucleons. The effects of the magnetic field modified Dirac sea are obtained through summation of scalar ($\sigma$, $\zeta$ and $\delta$) and vector ($\rho$ and $\omega$) tadpole diagrams in the evaluation of the nucleonic one-loop self energy functions, using the magnetized fermionic propagator within the chiral effective model. The Dirac sea contributes to the scalar densities of the nucleons. The contributions of both the magnetized Fermi and Dirac sea of nucleons are incorporated to the light quark and gluon condensates in terms of the scalar fields (equations \textbf{(8)-(12)}), through $\rho_{p,n}$ and $\rho^s_{p,n}$ of the nucleons within the chiral effective model. \\ 
 The present investigation is restricted to the cold nuclear matter effects. The light quark condensates ($\langle \bar{q}q \rangle$; $q=u,d$) and the scalar gluon condensate $(\langle\frac{\alpha_{s}}{\pi}G_{\mu\nu}^a G^{a\mu\nu}\rangle)$, thus obtained, are used to calculate the in-medium masses and other spectral parameters of the light axial-vector and vector mesons, by solving their respective finite energy sum rules [Eqs.\textbf{(42)-(47)}]. \\ 
 The values of the current quark mass and the running coupling constant, to be used in the present study are $m_u=4$ MeV, $m_d=7$ MeV and $\alpha_s=0.3551$ \cite{91, 100}. The vacuum masses of the light mesons are taken to be, $m_{A_1}=1230$ MeV; $m_{\rho}=770$ MeV and $m_{\omega}=783$ MeV \cite{pdg}. Using the vacuum masses, FESRs at vacuum (Eqs.\textbf{(35)-(37)} for $A_1$, and Eqs.\textbf{(38)-(40)} for $\rho$, $\omega$), are solved to obtain the coefficients $\kappa_j$ of the scalar four-quark condensates. The values obtained are, $\kappa_1=-2.204$ (for $A_1^0$); $\kappa_2=-2.485$ (for $A_1^{\pm}$); and $\kappa_3=-8.804$ (for $\rho^{\pm}$), $\kappa=7.237$ and $\kappa=7.789$ for the $\rho^0$ and $\omega$ mesons respectively. In the QCD sum rule study of the light vector mesons \cite{91}, the values of $\kappa_j=7.236, 7.788, -1.21$, were obtained for $j=\rho^0,\omega, \phi$ mesons, respectively, by solving their corresponding vacuum FESRs. For the axial-vector mesons, there is an extra pion pole contribution in the spectral density function, in addition to the $A_1$ meson pole. In our study, the pion decay constant, $f_{\pi}=93.3$ MeV and the pion mass, $m_{\pi}=139$ MeV \cite{papa} remain fixed. The vacuum values of the perturbative continuum threshold, $s_0$ and the resonance strength, $F$ (both in $GeV^2$) are thus found to be $1.266$ and $2.114$ (for $\rho$ meson state), $2.268$ and $2.755$ (for $A_1$ meson) and $1.3046$ and $0.2419$ (for $\omega$ meson), respectively. The value of the nuclear matter saturation density, $\rho_0=0.15\ fm^{-3}$ is used in our study \cite{papa}. 
 
\begin{figure}[h!]
    \centering
    \includegraphics[width=1.0\textwidth]{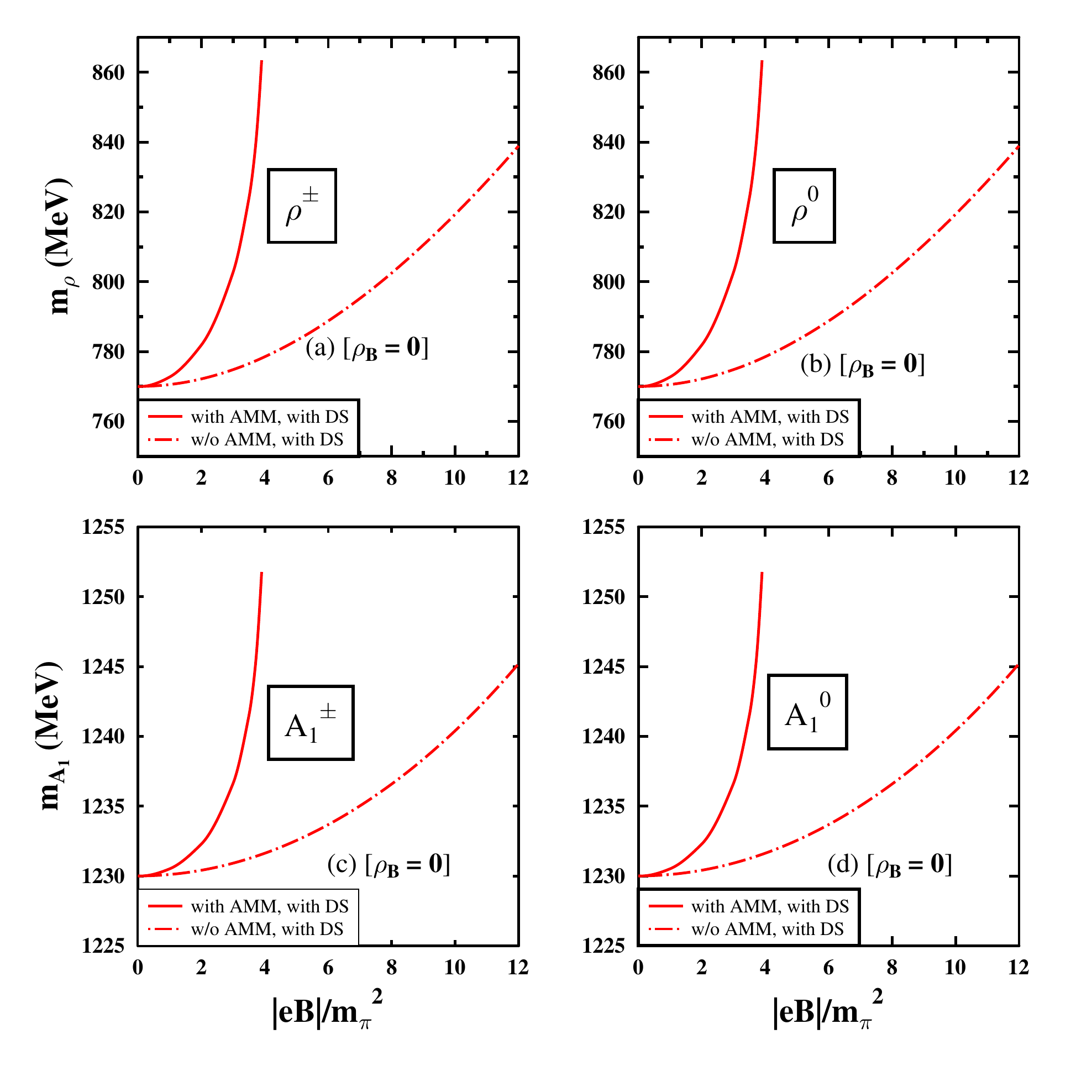}
    \vspace{-1.5cm}
    \caption{In-medium masses of $\rho^{\pm}$ (a), $\rho^{0}$ (b) and $A_1^{\pm}$ (c) and $A_1^{0}$ (d) states are plotted as functions of $|eB|$ (in units of $m_{\pi}^2$), at $\rho_B=0$. The effects of Dirac sea (DS) on the masses, are compared with the mass with no DS effect. The effects of the nucleons anomalous magnetic moments are considered and compared with the no AMM condition. } 
    \label{fig:0}
\end{figure}

 The figure \ref{fig:0} illustrates the in-medium masses of the $\rho$ and $A_1$ mesons with changing magnetic field, $|eB|$ in units of $m_{\pi}^2$, at zero baryon density $\rho_B=0$. The effects of nucleons' anomalous magnetic moments through the Dirac sea are compared to the case with zero AMM. At $\rho_B=0$, there is no additional contribution from the protons' Landau quantization. The magnetic field effect comes from the magnetized Dirac sea contribution only. At zero density, the light quark condensates increase with magnetic field, leading to the magnetic catalysis effect, both with and without the AMMs of the nucleons. The solutions of the scalar fields, at $\rho_B=0$ and for nonzero AMMs of the nucleons, are obtained up to $|eB|=3.9m_{\pi}^2$ in our present study. There is considerable amount of difference in the behavior of the light quark condensates through the scalar fields, due to the finite values of the anomalous magnetic moments of nucleons as compared to the zero AMM situation.

 \begin{figure}[h!]
    \centering
    \includegraphics[width=1.0\textwidth]{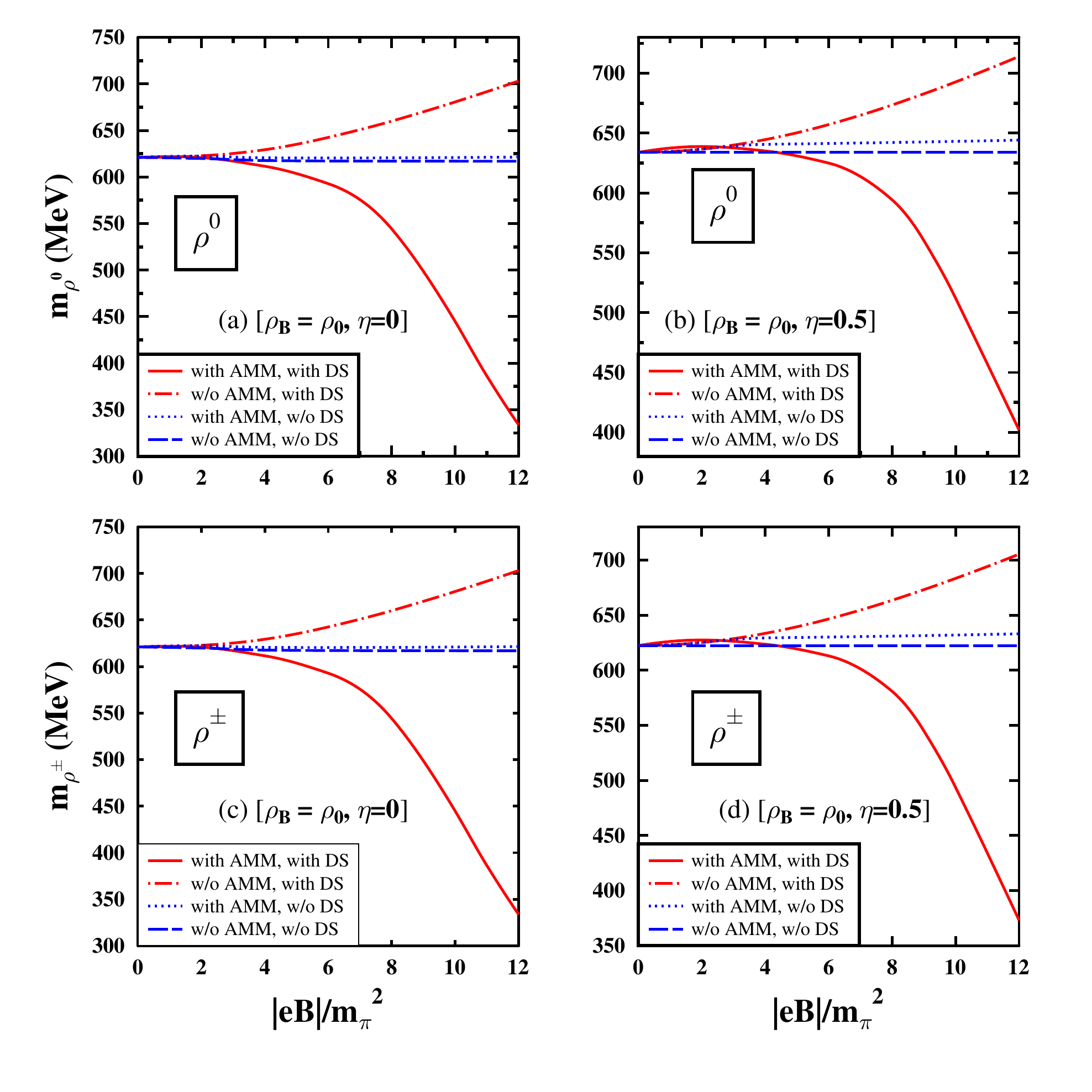}
    \vspace{-1.5cm}
    \caption{In-medium masses of $\rho^0$ [plots (a)-(b)] and $\rho^{\pm}$ [plots (c)-(d)] states are plotted as functions of $|eB|/m_{\pi}^2$, at $\rho_B=0,\ \rho_0$ and $\eta=0,\ 0.5$. The Dirac sea (DS) effects on the masses, are compared with the mass with no DS effect. The effects of the nucleons anomalous magnetic moments are considered and compared with the no AMM condition. } 
    \label{fig:1}
\end{figure}
\begin{figure}
    \centering
    \includegraphics[width=1.0\textwidth]{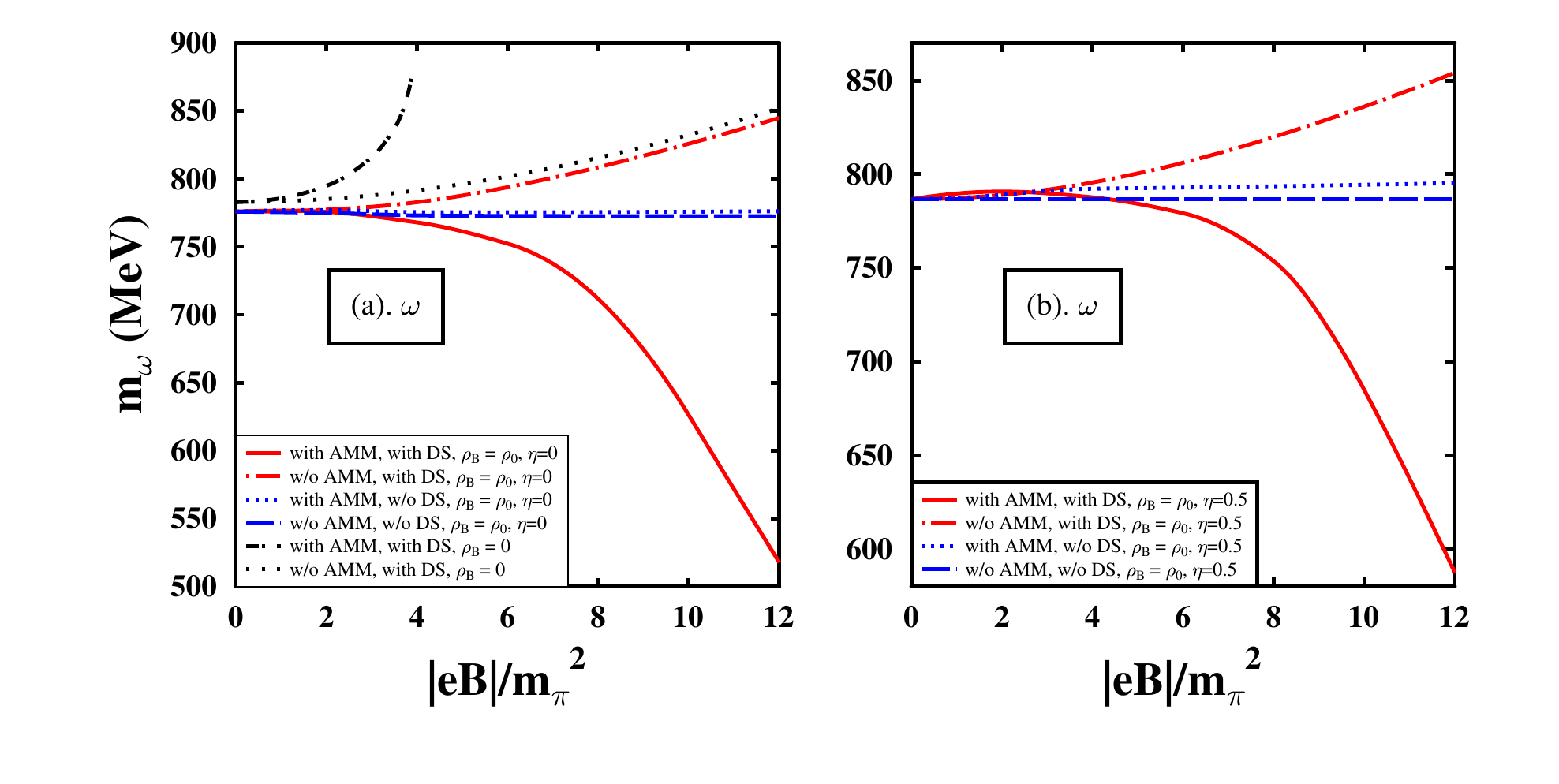}
    \vspace{-1.5cm}
    \caption{In-medium mass of $\omega$ state are plotted as functions of $|eB|/m_{\pi}^2$ at $\rho_B=0,\ \rho_0$ and $\eta=0,\ 0.5$. The Dirac sea (DS) effects on the masses, are compared with the mass with no DS effect. The effects of the nucleons AMMs are considered and compared with the no AMM condition.} 
    \label{fig:2}
\end{figure}
\begin{figure}
    \centering
    \includegraphics[width=1.0\textwidth]{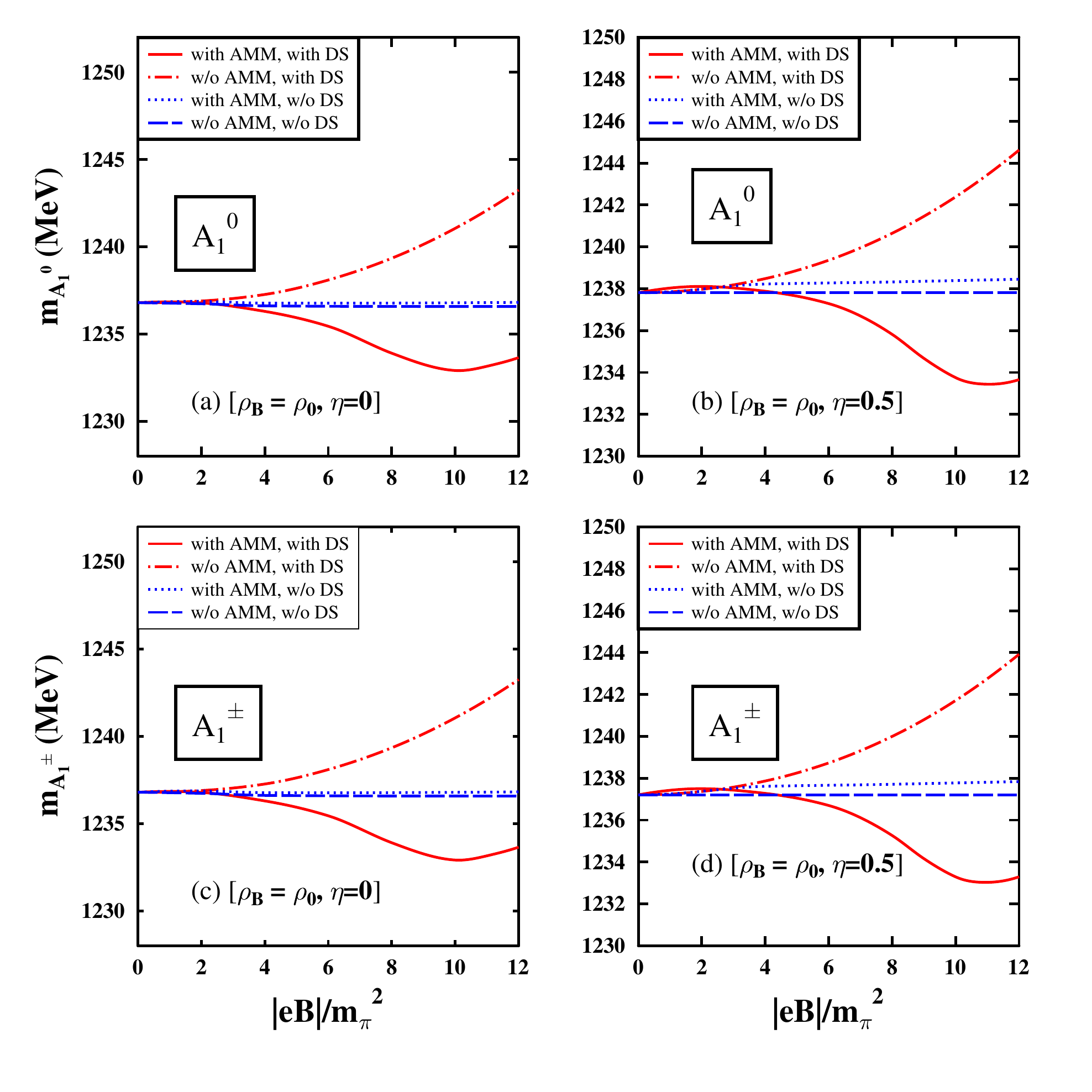}
    \vspace{-1.5cm}
    \caption{In-medium masses of $A_1^0$ [plots (a)-(b)] and $A_1^{\pm}$ [plots (c)-(d)] states are plotted as functions of $|eB|/m_{\pi}^2$ at $\rho_B=\rho_0$ and $\eta=0,\ 0.5$. The Dirac sea (DS) effects on the masses, are compared with the mass with no DS effect. The effects of the nucleons anomalous magnetic moments are considered and compared with the no AMM condition.} 
    \label{fig:3}
\end{figure}

 In figure \ref{fig:1}, the in-medium masses of the neutral [plots (a)-(b)], as well as the charged [plots (c)-(d)], $\rho$ meson are shown with variation in the magnetic fields, $|eB|/m_{\pi}^2$, at the nuclear matter saturation density, $\rho_0$ for symmetric ($\eta=0$) as well as asymmetric ($\eta=0.5$) nuclear matter. In the chiral $SU(3)$ model, incorporating the effects of the Dirac sea at finite magnetic field, the magnitudes of the light quark condensates decrease with increasing magnetic field, an effect called inverse magnetic catalysis, at $\rho_B=\rho_0$, for nonzero anomalous magnetic moments of the nucleons. The scalar gluon condensate has similar variation with $|eB|$, for nonzero AMMs, at the nuclear matter saturation density. At $\rho_B=0$, the behavior is opposite to the light quark condensates, indicate the nature of the scalar dilaton field, $\chi$ with $|eB|$ at zero density. The light quark condensates tend to rise with $|eB|$, at $\rho_0$, when the nucleons AMMs are taken to be zero, indicate a magnetic catalysis. Thus, the effects of the magnetized Dirac sea lead to inverse magnetic catalysis at  $\rho_B=\rho_0$, for nonzero AMMs of the nucleons. The magnetic catalysis is observed for zero AMM, at the nuclear matter saturation density. The effects of (inverse) magnetic catalysis are studied on the mass of the light vector and axial-vector mesons. In figure \ref{fig:1}, masses of the $\rho^{\pm}$ and $\rho^0$ mesons increase with magnetic field, accounting for the Dirac sea effects (denoted as DS), for zero AMMs of nucleons. There is observed to be rapid drop with $|eB|$, for nonzero AMMs of the nucleons. The effects of the magnetized Dirac sea on the $\rho$ meson masses are compared with the mass calculated without taking into account the DS effect, i.e., only Landau energy levels of protons contribute. The pattern in the mass variation are similar in symmetric and asymmetric matter, with slight difference in the magnitudes. Figure \ref{fig:2} shows the mass of the $\omega$ meson as a function of $|eB|/m_{\pi}^2$, at $\rho_B=0$ (in (a)), and $\rho_0$ for $\eta=0$ (in (a)) and $\eta=0.5$ (in (b)). Similar mass variation is observed for the $\omega$ meson like the $\rho$ meson case. The anomalous magnetic moments of the nucleons play important role through the magnetized Dirac sea contribution, both at zero and finite density matter. In figure \ref{fig:3}, mass of the neutral $A_1^0$ [plots (a)-(b)], and the charged $A_1^{\pm}$ [plots (c)-(d)] axial-vector mesons, are plotted as functions of $|eB|/m_{\pi}^2$, at $\rho_0$ and $\eta=0,\ 0.5$. For the axial-vector meson, the in-medium masses at $\rho_B=\rho_0$ are slightly larger as compared to the vacuum mass, which is opposite in nature with its chiral partner, $\rho$ meson. This is due to the sign of the scattering term considered in the finite energy sum rules (eq.(42) for $A_1$ meson states). Although masses of the $A_1$ mesons decrease first and then slightly rise with increasing magnetic field, at $\rho_B=\rho_0$ and for nonzero anomalous magnetic moments of the nucleons. This is when considering the Dirac sea effects. There are observed to be no changes in mass, when DS effect is not considered. For the zero AMM case, with DS effects, masses are seen to increase with $eB$, just like the $\rho$ meson.
 
\begin{figure}[h!]
    \centering
    \includegraphics[width=1.0\textwidth]{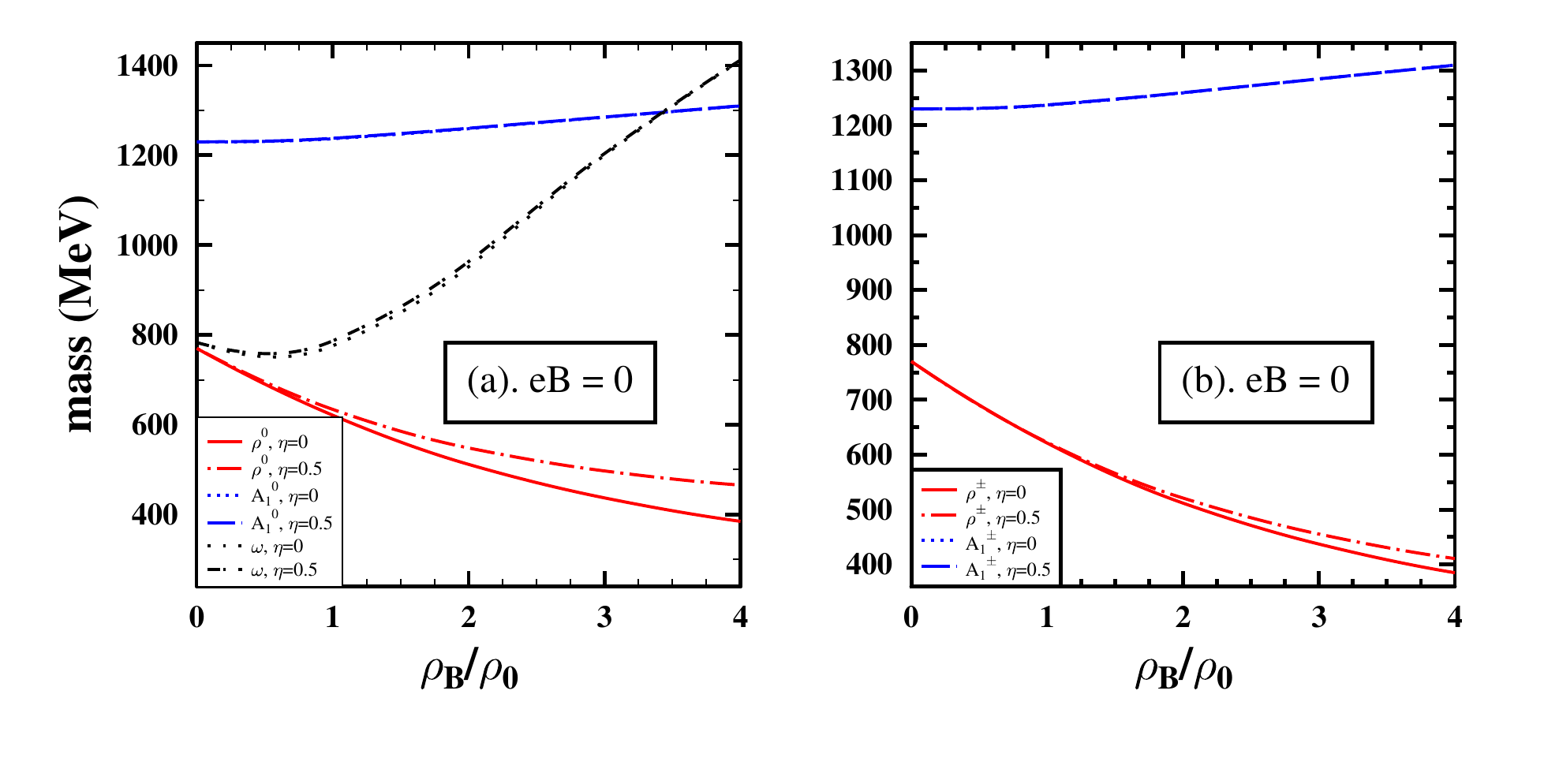}
    \vspace{-1.5cm}
    \caption{Mass of the neutral mesons, $\rho^0,\ \omega,\ A_1^0 $ [plot (a)] and charged particle states $\rho^{\pm},\ A_1^{\pm}$ [plot (b)] are plotted as functions of the relative baryon density, $\rho_B/\rho_0$, at zero magnetic field ($|eB|=0$) and for the isospin asymmetry parameter, $\eta=0,\ 0.5$.} 
    \label{fig:4}
\end{figure}

\begin{figure}[h!]
    \centering
    \includegraphics[width=1.0\textwidth]{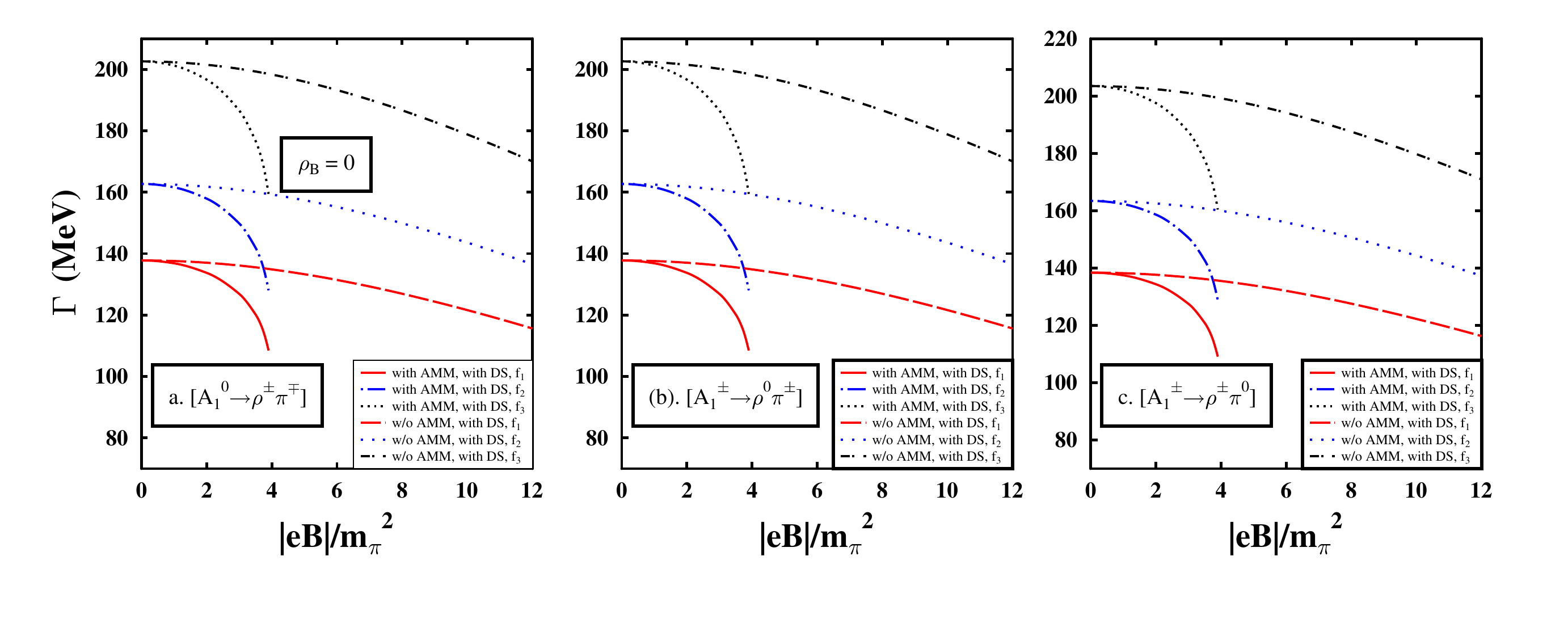}
    \vspace{-1.5cm}
    \caption{ The hadronic partial decay widths for the $A_1^0\rightarrow \rho^{\pm}\pi^{\mp}$ (a), $A_1^{\pm}\rightarrow \rho^{0}\pi^{\pm}$ (b) and $A_1^{\pm}\rightarrow \rho^{\pm}\pi^{0}$ (c) channels, are plotted as functions of $|eB|/m_{\pi}^2$ at $\rho_B=0$, accounting for the Dirac sea effects. The effects of nucleons anomalous magnetic moments are compared with the case when AMM is not considered. The decay widths are shown for all three set of coupling constant, $f=f_1,\ f_2,\ f_3$.} 
    \label{fig:5}
\end{figure}

\begin{figure}[h!]
    \centering
    \includegraphics[width=1.0\textwidth]{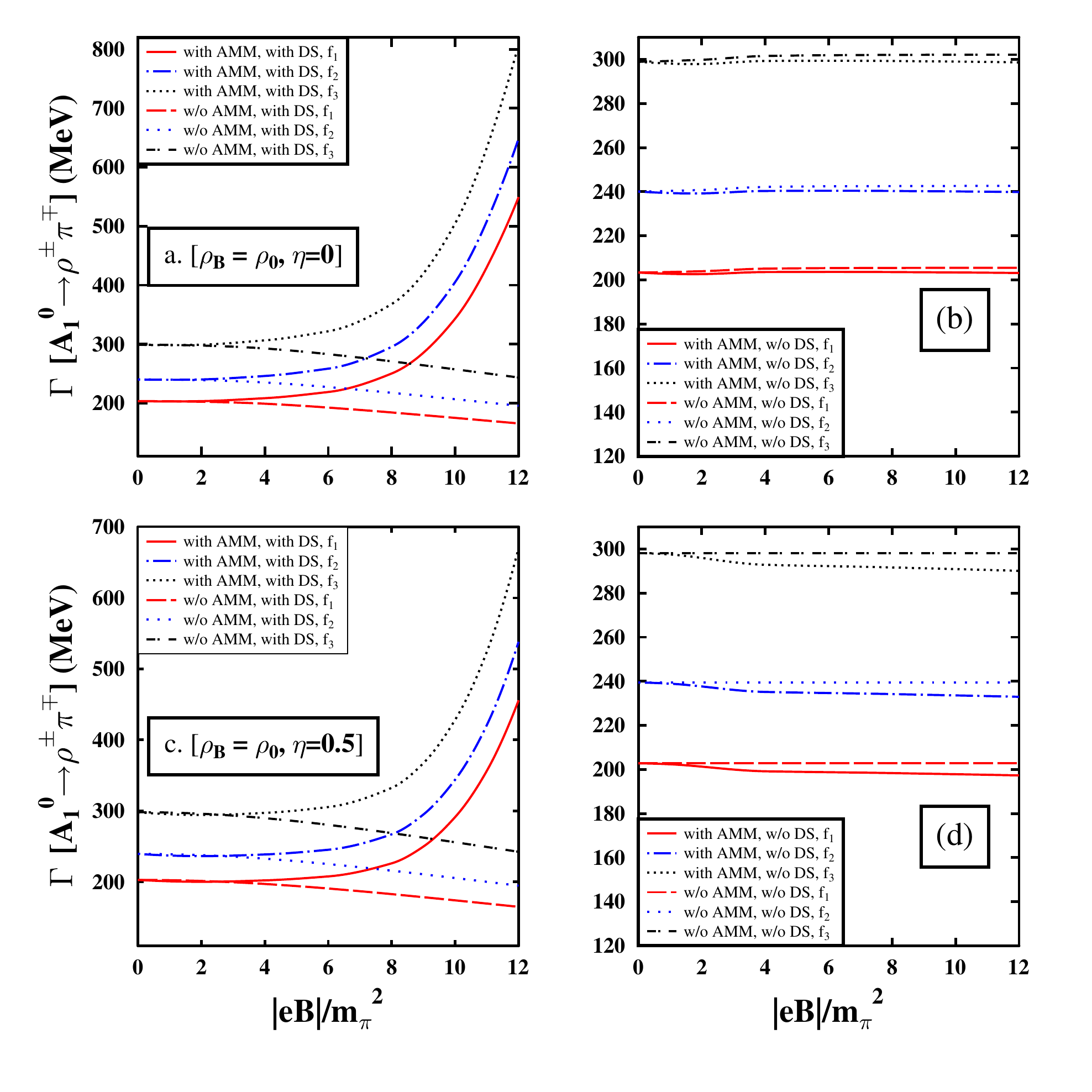}
    \vspace{-1.5cm}
    \caption{In-medium, partial decay widths for the $A_1^0\rightarrow \rho^{\pm}\pi^{\mp}$ mode, are plotted as functions of $|eB|/m_{\pi}^2$ at $\rho_B=\rho_0$, and for $\eta=0$ [in (a)-(b)], $\eta=0.5$ [in (c)-(d)]. The effects of Dirac sea (DS) [in (a) and (c)] are compared with the case when there is no Dirac sea effect [(b) and (d)]. Comparison on the basis of nucleons AMMs are shown.} 
    \label{fig:6}
\end{figure}

\begin{figure}[h!]
    \centering
    \includegraphics[width=1.0\textwidth]{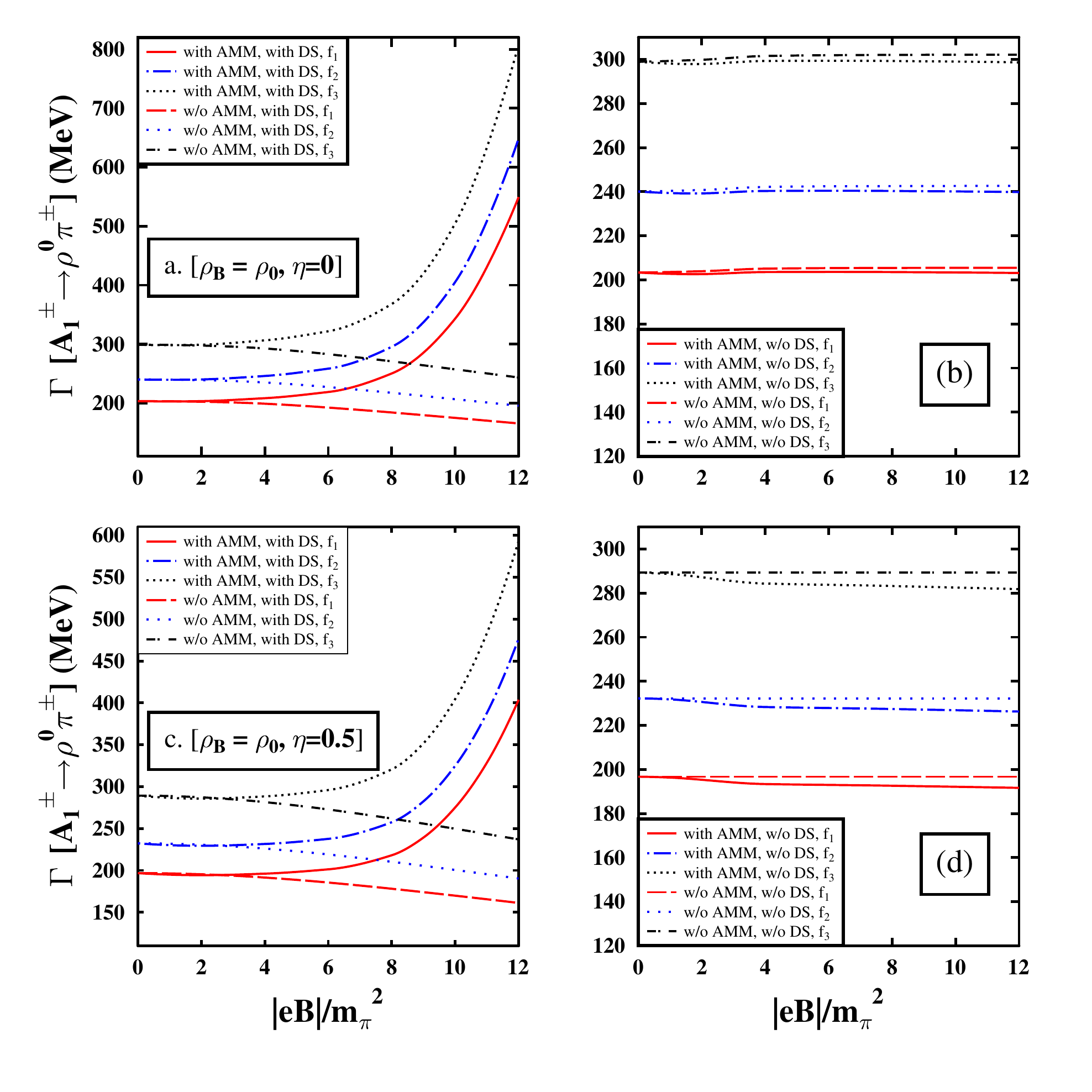}
    \vspace{-1.5cm}
    \caption{In-medium partial decay widths for the $A_1^{\pm}\rightarrow \rho^{0}\pi^{\pm}$ mode, are plotted as functions of $|eB|/m_{\pi}^2$ at $\rho_B=\rho_0$, and for $\eta=0$ [plots (a)-(b)], $\eta=0.5$ [plots (c)-(d)]. The effects of Dirac sea (DS) [in (a) and (c)] are compared with the case when there is no Dirac sea effect [(b) and (d)]. Comparison on the basis of nucleons AMMs are shown.} 
    \label{fig:7}
\end{figure}

\begin{figure}[h!]
    \centering
    \includegraphics[width=1.0\textwidth]{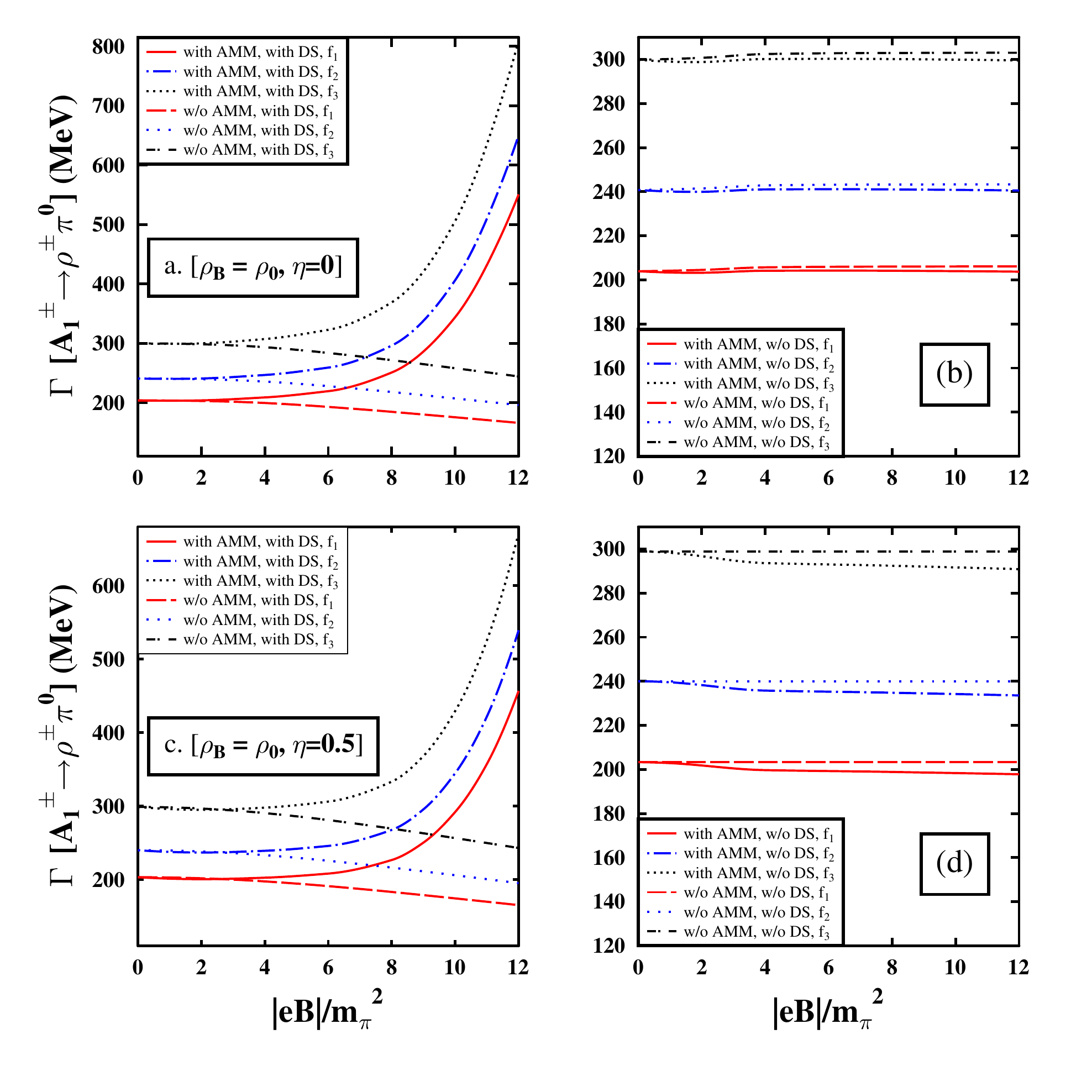}
    \vspace{-1.5cm}
    \caption{The in-medium partial decay widths for the $A_1^{\pm}\rightarrow \rho^{\pm}\pi^{0}$ mode, are plotted as functions of the magnetic field, $|eB|/m_{\pi}^2$ at $\rho_B=\rho_0$, and for $\eta=0$ [plots (a)-(b)], $\eta=0.5$ [plots (c)-(d)]. The effects of Dirac sea (DS) [in (a) and (c)] are compared with the case when there is no Dirac sea effect [(b) and (d)]. Comparison on the basis of nucleons AMMs are shown.} 
    \label{fig:8}
\end{figure}

\begin{figure}[h!]
    \centering
    \includegraphics[width=1.0\textwidth]{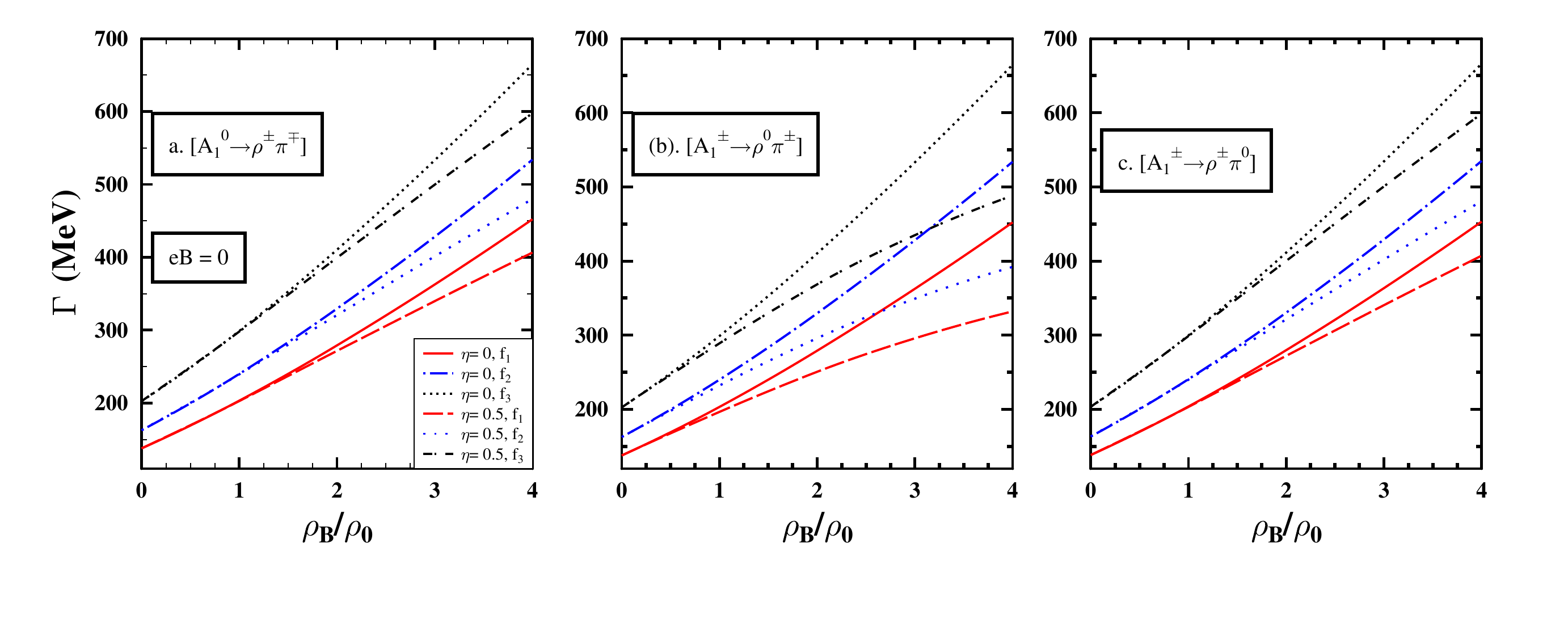}
    \caption{The hadronic partial decay widths for the $A_1^0\rightarrow \rho^{\pm}\pi^{\mp}$ [plot (a)], $A_1^{\pm}\rightarrow \rho^{0}\pi^{\pm}$ [plot (b)] and $A_1^{\pm}\rightarrow \rho^{\pm}\pi^{0}$ [plot (c)] modes of decay, are plotted as functions of the relative baryon density, $\rho_B/\rho_0$, at $|eB|/m_{\pi}^2=0$, for all three set of coupling constant, $f=f_1,\ f_2,\ f_3$.} 
    \label{fig:9}
\end{figure}

In Fig.\ref{fig:4}, masses are plotted as functions of the baryon density, $\rho_B$ (in units of $\rho_0$), at $|eB|=0$. Masses are shown in plot.(a) for the neutral states, at isospin symmetric ($\eta=0$) as well as the asymmetric ($\eta=0.5$) nuclear matter and plot.(b) for the charged particle states. In-medium masses of $\rho$ meson decrease with increasing density, that of $\omega$ meson decrease first, then rise sharply with density. The mass of the light axial-vector meson, $A_1$, which is the chiral partner of the light vector meson, $\rho$, increases very slightly with density.
The in-medium partial decay widths for the hadronic decay modes, $A_1^0\rightarrow \rho^{\pm}\pi^{\mp}$ [plot (a)], $A_1^{\pm}\rightarrow \rho^{0}\pi^{\pm}$ [plot (b)] and $A_1^{\pm}\rightarrow \rho^{\pm}\pi^{0}$ [plot (c)] are plotted in fig.\ref{fig:5}, as functions of the magnetic field, $|eB|/m_{\pi}^2$ at $\rho_B=0$. This is when the effects of the Dirac sea are taken into account on the masses of the initial and final states particles, at $\rho_B=0$. At zero density, the effects of the resulting magnetic catalysis on the masses lead to a drop in the decay widths with increasing magnetic field for all three channels. By using the experimentally fitted values of the relevant decay modes as mentioned in table.(4) of Ref.\cite{roca70}, three set of values for the Lagrangian parameters are obtained. The values of $\tilde{f}=$ $1270 \ (f_1)$, $1380\ (f_2)$ and $1540\ (f_3)$ in MeV, thus obtained are used in the present study to calculate the desired decay widths. 
 The decay width for the $A_1\rightarrow\rho\pi$ mode, is calculated using Eq.(57), for all three parameters separately and the masses of the corresponding meson states as have already discussed before in this section. The mass of the neutral and charged pions to be used here are $134.97$ MeV and $139.57$ MeV respectively \cite{pdg}. The vacuum decay widths for $(A_1^0\rightarrow\rho^{\pm}\pi^{\mp})/(A_1^{\pm}\rightarrow\rho^{0}\pi^{\pm})/(A_1^{\pm}\rightarrow\rho^{\pm}\pi^{0})$ channels are thus calculated to be $137.776/137.7756/138.3684$ MeV, $162.6763/162.6758/163.3758$ MeV and $202.5851/202.5845/203.4562$ MeV, for $\tilde{f}=f_1, f_2$ and $f_3$ respectively. These values are very close to the central average value of $210$ MeV (corresponding to the $50\%$ branching ratio) for $A_1\rightarrow\rho\pi$ channel, as given in \cite{roca70}. The phenomenological interaction Lagrangian contains no term involving $A_1^{0}\rho^0\pi^0$ vertex, thus there is no decay for $A_1^0\rightarrow \rho^0\pi^0$ mode. The authors of CLEO Collaboration \cite{tao} have been reported a branching ratio of about $60\%$ for the intermediate $\rho\pi$  ($S$-wave) state of $A_1^{-}$ meson decaying into $\pi^- \pi^0 \pi^0$ relative to the total $A_1^-\rightarrow \pi^- \pi^0 \pi^0$ decay. The works of \cite{chen91} showed $\rho\pi$ ($S$-wave) to be the dominant decay channel for the ground state of $A_1$ meson in its class of axial-vector mesons with quantum numbers $(I^G(J^{PC})=1^-(1^{++}))$. They have analyzed the mass spectra of all the possible light axial-vector meson states by Regge trajectory analysis and have applied a light quark pair creation model \cite{micu} to calculate their OZI-allowed strong decays. In figs.[\ref{fig:6}-\ref{fig:8}], the in-medium partial decay widths for $(A_1^0\rightarrow\rho^{\pm}\pi^{\mp}),\ (A_1^{\pm}\rightarrow\rho^{0}\pi^{\pm}),\ (A_1^{\pm}\rightarrow\rho^{\pm}\pi^{0})$, channels respectively, are plotted as functions of the magnetic field, at $\rho_B=\rho_0$ and $\eta=0,\ 0.5$. 
The plots (a) and (c) in these figures show the effects of Dirac sea (denoted as DS), on the decay widths through their in-medium masses, both with and without considering the anomalous magnetic moments of the nucleons. There is almost no change in the decay widths with rising magnetic field, when DS effect is not considered on the mass [plots (b) and (d)]. Finally, fig.\ref{fig:9} illustrates the variation of the in-medium decay widths as functions of the relative baryon density $\rho_B/\rho_0$, at zero magnetic field ($|eB|=0$). Since the masses of the decay product, $\rho$, decrease as the density increases, with almost no changes in the parent particle masses ($A_1$ state), the decay widths are observed to rise with density. In our present study, no in-medium effect considered on the mass of the pseudoscalar meson ($\pi$). 
\section{Summary}
\label{sec6}
In the summary, we have investigated the in-medium masses and hadronic decay widths of the light vector and axial-vector mesons in a magnetized nuclear matter, taking into account the effects of the Dirac sea. The medium modifications of masses are calculated within the sum rule framework for the $\rho^{0, \pm}$, $\omega$ and $A_1^{0, \pm}$ mesons, using the finite energy sum rules. The values of the other spectral parameters ($F,\ s_0$) are modified by the in-medium effects. The in-medium masses are obtained in terms of the medium modified light quark (up to the scalar four-quark condensate) and the scalar gluon condensates. These condensates are calculated from the scalar fields within the chiral $SU(3)$ model. The effects of magnetic field due to the Landau energy levels of protons, anomalous magnetic moments of the nucleons, and the magnetized Dirac sea of nucleons, are studied on the in-medium properties of light mesons. Dirac sea effect contributes to the scalar densities of the nucleons. The (reduction) increment in the values of the light quark condensates with increasing magnetic field, give rise to the (inverse) magnetic catalysis effect. A phenomenological Lagrangian approach is adopted to account for the $av\phi$ interaction vertices, in order to calculate the decay width of an axial-vector meson going to a vector and a pseudoscalar meson, i.e., $A_1\rightarrow \rho\pi$. As the in-medium masses of the parent ($A_1$) and daughter ($\rho$) particles (decrease) increase with magnetic field due to the (inverse) magnetic catalysis, the corresponding partial decay widths for the possible modes of $A_1\rightarrow \rho\pi$ channel, (rise) drop with rising magnetic field, accounting for the Dirac sea effect. There is almost no medium effects on the masses and decay widths, with the variation in magnetic field, when the DS effect is not considered. Thus, the appreciable changes in the in-medium properties of the $\rho$, $\omega$ and $A_1$ mesons, due to the Dirac sea effect, at finite magnetic field, may affect the experimental observables in the peripheral, ultra relativistic, heavy-ion collision experiments, where produced magnetic field is huge. 

\acknowledgements
Amruta Mishra acknowledges financial support from Department of Science and Technology (DST), Government of India (project no. CRG/2018/002226).

\end{document}